\documentclass[runningheads]{llncs}
\usepackage{textgreek}
\usepackage{graphicx}
\begin{document}
\title{Wearable Haptic Device for Individuals with Congenital Absence of Proprioception}
\titlerunning{Wearable Haptic Device for Proprioceptive Feedback}
%
\author{Sreela Kodali\inst{1}\and
Allison M. Okamura\inst{1}\and
Thomas C. Bulea\inst{2}\and\\
Alexander T. Chesler\inst{2}\and
Carsten G. Bönnemann\inst{2}}
%
\authorrunning{S. Kodali et al.}
%
\institute{Stanford University, Stanford CA 94305, USA\\
\email{\{kodali, aokamura\}@stanford.edu} \and
National Institutes of Health, Bethesda, MD 20892, USA\\
\email{\{thomas.bulea, alexander.chesler, carsten.bonnemann\}@nih.gov}}


%
\maketitle              
\begin{abstract}
A rare genetic condition, PIEZO2 loss of function (LOF) is characterized by absence of proprioception and light touch, which makes functional tasks (e.g., walking, manipulation) difficult. There are no pharmacological treatments or assistive technologies available for individuals with PIEZO2-LOF.  We propose a sensory substitution device that communicates proprioceptive feedback via detectable haptic stimuli. We created a wearable prototype that maps measurements of elbow movement to deep pressure applied to the forearm. The prototype applies up to 18 N, includes an embedded force sensor, and is programmable to allow for various angle-to-pressure mappings. 
Future work includes comparing proprioceptive acuity and movement ability with and without the device in healthy and PIEZO2-LOF individuals, developing low-profile devices using soft robotics, providing sensory substitution for multiple joints simultaneously, and encoding additional aspects of joint dynamics.

\end{abstract}
\section{Introduction}
Proprioception can be considered our ``sixth sense.'' It provides continuous information on body position and movement vital to motor control and coordination, balance, muscle tone, postural reflexes, and skeletal alignment. Many neuromuscular disorders arise from dysfunction of motor efferents, but a deficiency of afferent proprioceptive sensory input is another, often overlooked, cause of impairment that can severely impact motor function, even when strength is preserved. Proprioception in humans is entirely dependent on the non-redundant mechanosensor PIEZO2; individuals with recessive PIEZO2-LOF show complete 
congenital absence of proprioception leading to motor and functional impairment \cite{ref_nEnglJMed_Piezo2}. 
Individuals with PIEZO2-LOF also lack vibratory sense and discriminatory touch perception specifically
on glabrous skin, although deep pressure, temperature, and some pain sensation is preserved. \cite{ref_innocuous,ref_sensor,ref_nEnglJMed_Piezo2,ref_ScienceTranslational_Piezo2}. 
No therapeutic or assistive technology options currently exist for individuals with PIEZO2-LOF. Our goal is to design and test a wearable haptic device that enables proprioceptive feedback using preserved sensory input modalities and evaluate its efficacy to enable intuitive control of limb movement in individuals with PIEZO2-LOF. 

\begin{figure}[t]
\includegraphics[width=\textwidth]{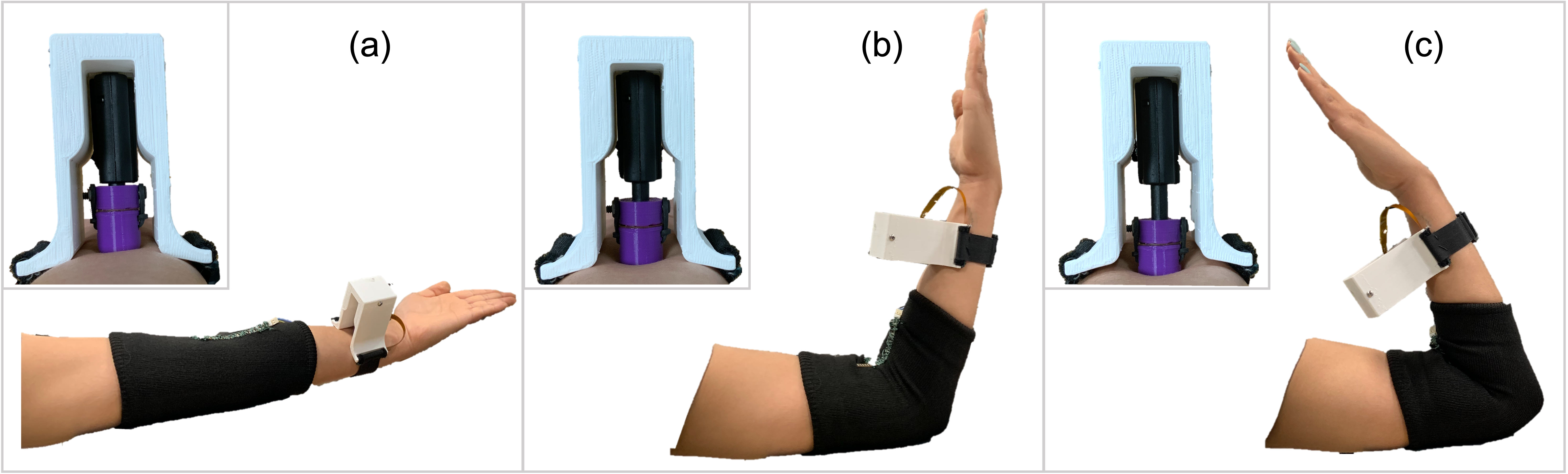}
\caption{Prototype device with elbow angle measurement and tactor mounted on the forearm. As elbow angle becomes more acute, the tactor provides increased pressure.} \label{fig1}
\end{figure}


\section{Prototype and Preliminary Results}
Because individuals with PIEZO2-LOF have intact deep pressure sensation, we designed a prototype device that transduces elbow angle information to deep pressure stimuli applied to the forearm. The haptic feedback component consists of a linear actuator and cylindrical tactor housed in a plastic enclosure worn flush against the surface of the arm. The device is fastened in place with wide hook-and-loop straps. As the actuator extends its position, the tactor presses directly on the skin and applies a deep pressure stimulus. We used an Actuonix PQ12-P micro linear actuator because of its built-in position sensor, high output force (18 N), and reliable control. We embedded a low-profile SingleTact capacitive force sensor (diameter 15 mm, force range 45 N) in the tactor to measure the applied pressure in real time.

We measured elbow angle with a resistive flex sensor (20-40 k\textOmega \hspace{0.05cm} range in series with 42 k\textOmega \hspace{0.05cm} resistor as a voltage divider) sewn into a fabric sleeve. A microcontroller maps the elbow angle to a pressure stimulus, sends a position to the linear actuator, and records the force and actuator position. The real-time feedback and precise control of the actuator will allow us to explore different stimulus patterns. Figure 1 depicts the device behavior when the 
angle-to-pressure mapping is a constant 0.123 mm/deg. When the elbow is fully extended and arm is outstretched, the actuator is retracted and applies no pressure (Fig.~1a). When the elbow and arm are bent, the actuator is proportionately extended and applies pressure (Fig.~1b). When the elbow is fully flexed, the actuator is maximally extended and applies the most pressure (Fig.~1c).

Figure 2 shows measurements of elbow angle, actuator position, and force during repetitions of the movement shown in Fig.~1. Our results confirmed the device functioned as expected. 
The actuator position and force are affected by noise in the flex sensor and friction in the actuator.

\noindent In addition, to verify force sensor functionality, we measured the force-displacement relationship of the device when mounted on two locations on the body: the ventral side of the forearm and the dorsal side of the hand. Because the hand is locally stiffer than the forearm, we expected that the forces would be larger for a given displacement at the hand compared to the forearm. The measurements corroborated this; we identified the arm and hand surfaces with stiffness values of 465.6 and 8115.4 N/m, respectively.

\begin{figure}[t]
\includegraphics[width=\textwidth]{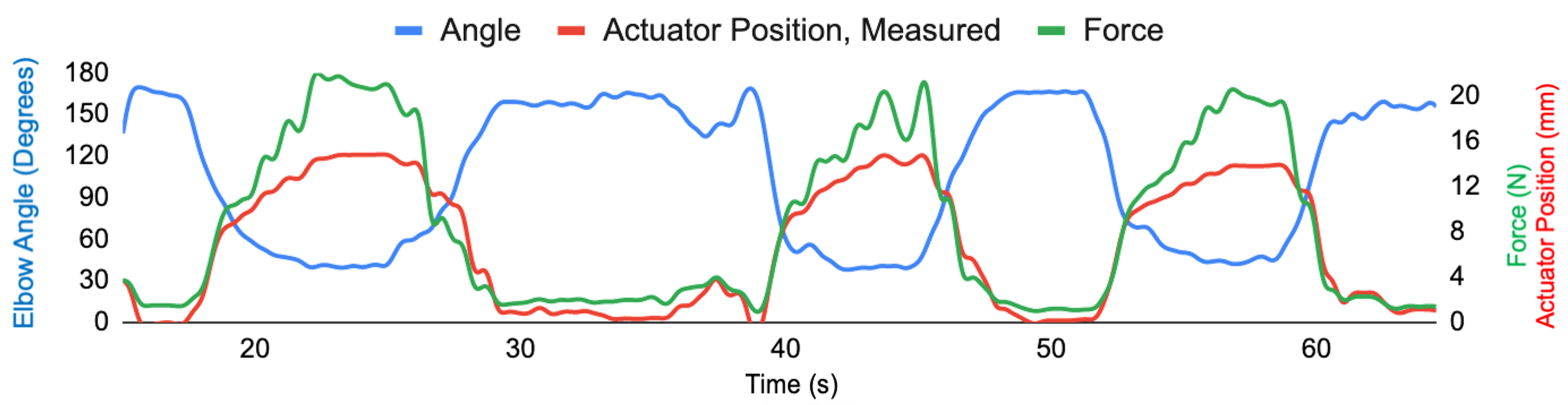}
\caption{Filtered measurements of elbow angle, actuator position, and force recorded during use of the prototype. Acute angles are flexion and 180 degrees is full extension. } \label{fig2}
\end{figure}

\section{Conclusions and Future Work}
We created a wearable sensory substitution device for eventual use by individuals with PIEZO2-LOF to convey proprioceptive feedback via haptic stimuli. Characterizing the sensory input of PIEZO2-LOF patients is  ongoing, so we designed our prototype device to be programmable and cover a wide force range. The device successfully mapped elbow joint angle to single site deep pressure stimuli on the forearm. 
For our next steps, we will replace the resistive flex sensor with a robust capacitive sensor for angle measurement and use the programmable device to evaluate different designs (e.g. stimulus patterns, tactor sizes, mappings) and quantify their impact on PIEZO2-LOF patients' proprioceptive acuity. We will also explore the use of multiple deep pressure units, spatiotemporal stimuli patterns, and new designs based on soft robotics techniques. If the elbow device is effective, we will proceed to encode multiple degrees of freedom of joint movement for both the upper and lower limbs. 

%
%

%
%


\end{document}